\documentstyle[aps,prb,twocolumn]{revtex}

\newcommand\be{\begin{equation}}
\newcommand\ee{\end{equation}}

\input epsf
\epsfclipon

\begin{document}

\wideabs{

\title{Strain in heteroepitaxial growth}
\author{E.~Somfai and L.~M.~Sander}
\address{Department of Physics, The University of Michigan,
	Ann Arbor MI 48109-1120}
\date{\today}
\maketitle

\begin{abstract}
We use atomistic simulations with an empirical potential
(EAM) to study the elastic effects of heteroepitaxial islands on adatom
diffusion. We measure the diffusion barrier on pure stressed substrate and
near a misfit island, as well as the detachment barrier from islands of
different size.
\end{abstract}

} 

\section{Introduction}

Heteroepitaxy is the growth by deposition of one material on another.  Since
the two materials are different, stress can be generated, and it leads to a
number of interesting and important phenomena.  In this paper we consider
various effects due to this strain by using the embedded atom method (EAM)
\cite{eam}.

We consider the case where the adatoms are ``larger'' than the substrate
atoms.  Then if the adlayers are \emph{pseudomorphic} (follow the periodic
order of the substrate without dislocations), they have to be compressed.
The size difference is not necessarily the only source of stress:  the
compressive stress of few monolayers of Ag on Pt(111) five times larger than
expected from the size difference \cite{grossmann} presumably due to charge
transfer.  Since the elastic energy of the stressed layer is proportional to
its height, the excess elastic energy can, in the course of growth, overcome
the barrier of creating a dislocation network for relaxation.  Thus
pseudomorphic growth cannot be stable for large thickness.  

However
dislocated growth is not the only possibility: the adatoms can form
three-dimensional islands \cite{Snyder} instead of normal layer-by-layer
growth by merging of two-dimensional islands. The energetic reason
for this is that three-dimensional islands can relax: the lattice 
constant in the majority of an island
can be close to its bulk value, and only the bottom of the island is
stressed significantly.  In case of Volmer--Weber growth\cite{volmer}, the
islands nucleate on the substrate, while in the
Stranski--Krastanov case \cite{stranski} the first few layers grow
epitaxially (\emph{wetting layers}), and then islands form.

Three-dimensional islands are important for practical applications, as they
are a good candidate for lateral electron confinement.  Certain semiconductor
systems (e.g., InAs on GaAs) develop pseudomorphic three-dimensional
structure with a relatively narrow size distribution for the islands
\cite{notzel,Kobayashi}.  Since this ordering
takes place spontaneously during epitaxial growth the islands are called
\emph{self-organized quantum dots}.  The uniformity of quantum dots and, in
particular, the possible narrowing of the island size distribution due to
strain, is an important technological issue, and considerable effort has gone
into measuring \cite{Kobayashi,ebiko} and modeling
\cite{Williams,orr,dobbs1,dobbs2,Zangwill} this effect.

A simplified atomic level simulation of strained epitaxial systems has been
done by Orr \textit{et al.}\cite{orr} using the dynamic Monte--Carlo method
in one dimension. Surface particles were allowed to hop to neighboring
sites. The hopping probability depended on both the bond and the local
strain.  The strain was found by local relaxation after each motion and
global relaxation after fixed number of timesteps.  The elastic lattice was
modeled with harmonic forces between nearest and next-nearest neighbors.
Qualitative effects that are observed in experiment are found in this
treatment, e.g.  the formation of three-dimensional islands whose size
distribution depends on growth rate.  However, the treatment is very
schematic, and contains free parameters.

More realistic theories fall in two classes. Empirical approaches
\cite{Williams,dobbs1,dobbs2,Zangwill} take as 
input certain presumed effects on growth in heteroepitaxy
such as the tendency to detach from large two-dimensional islands
due to the build-up of strain, the effects of strain on adatom diffusion,
and the probability of conversion of a two-dimensional island
to a three-dimensional one. However, these parameters are
based on a combination of fitting and classical elastic theory with
no way to estimate the size of various effects for a real material. 

Schroeder and Wolf \cite{schroeder} studied the effect of strain on
surface diffusion for a Lennard-Jones lattice. This treatment can
potentially estimate all the relevant effects. However, the
Lennard-Jones potential is quite different from that found in the
substances of interest.  They observed that the activation barrier is
a linear function of strain over a wide range:
compressive strain enhances diffusion, while tensile strain hinders
it.  The strain changed mostly the energy of the saddle point, the
stable sites were not much affected.  The strain field of a
coherent two-dimensional island is not uniform (the edges are more
relaxed than the center), therefore this is reflected on the diffusion
of adatoms on top of the island.

In this paper we use more realistic potentials than 
Schroeder and Wolf \cite{schroeder},
namely the EAM approximation.  Unfortunately, this method is
appropriate for metals, not semiconductors, which are the materials of
most technological interest.  However, these potentials are probably
more reliable than the various phenomenological interactions which have
been proposed for semiconductors and serve to give perspective on the
effects which can be important in growth.  Also, this is a
computationally intensive approach, and can only treat one island at a
time.  Thus, we can only calculate parameters which can eventually be
inputs into an empirical theory.  To this end we investigate effect of
strain on the diffusion barrier of adatoms, the detachment energy from
a two-dimensional island, and the energy landscape for diffusion
nearby.

\section{Simulations}
In our simulations we use a substrate of slab geometry, periodic in the
lateral directions, with an open surface at the top bounded by a frozen
lattice below. The atoms of the substrate and the adlayers or adatoms
were allowed to relax according to the potential described below. We did
not introduce dislocations in the substrate. The relaxation was achieved by
using conjugate gradient methods.

It is necessary to have the elastic part of the substrate as deep as
wide, because the elastic effects penetrate roughly
isotropically\cite{schroeder}.  If the lattice was shallower then the
deformation field would be cut off and we would lose long range
effects.  This restriction has severe consequences on the lattice
sizes that are computationally tractable.

For an interatomic potential, we used the embedded atom method (EAM)\cite{eam}
which is believed to give a good representation of transition metals.
The form of the potential is
\be
  E_{\rm tot} = \frac{1}{2}\sum_i\sum_{j(\not=i)} \phi^{(ij)}(R_{ij})
  + \sum_i F^{(i)}(\rho^{\rm host}_i)
\ee
The pair potential part, $\phi^{(ij)}(R)$ is attributed to electrostatic
interactions, while the embedding function $F^{(i)}(\rho^{\rm host})$ is
interpreted as the interaction of the ion cores with the free electrons.
The explicit form of the functions used in our simulation are given in
Ref.~\onlinecite{eam}.

This pseudopotential provides reasonable values for many bulk
properties.  Whether it is appropriate for surface simulations is not
as clear for various reasons.  Nevertheless, the EAM is more realistic
approach than pair potentials such as Lennard-Jones, and
computationally tractable for the necessary system sizes as opposed to
first principle calculations.  We selected the Ag/Ni system based on
its large misfit (16\% compression of the Ag adlayer) from the
elements available with the EAM potential. This way we could achieve
significant stress in the islands on a substrate of relatively small,
computationally tractable size: $32^3$ in the following calculations.

\section{Results} 
\begin{figure} 
    \epsfxsize=8.6cm 
    \epsfbox{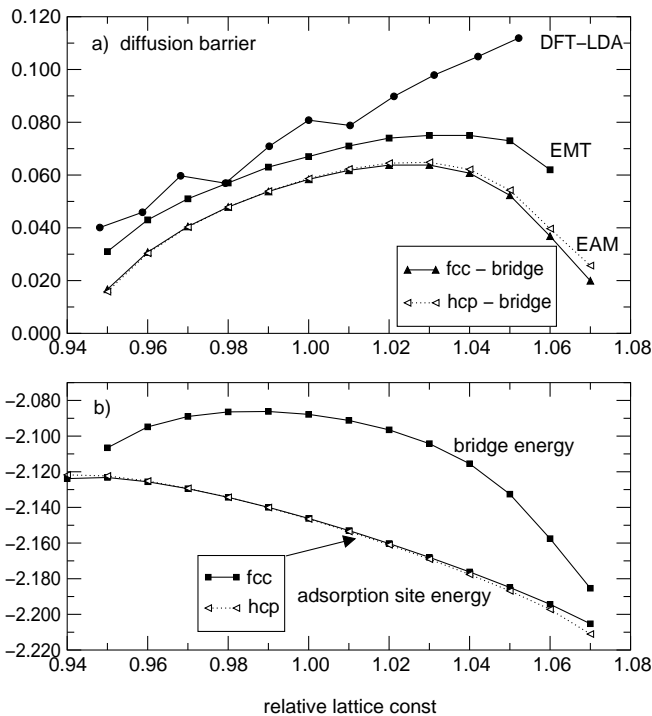}
    \bigskip
    \caption[Diffusion barrier of Ag adatom on stressed Ag(111) substrate]{
    Diffusion barrier of Ag adatom on stressed Ag(111) substrate.
    a) Comparison of the diffusion barrier obtained by density-functional
    theory (circles, from Ref.~\onlinecite{Ratsch}), effective medium theory
    (squares, from Ref.~\onlinecite{brune}) and our calculations with EAM
    potential (triangles).
    The barrier is plotted
    against the ratio of the stressed and the equilibrium lattice constant.
    b) The effect of strain on the bridge energy and the adsorption site
    energy of the same system (using EAM potential).  Note that around zero
    stress, the bridge energy is relatively constant, while the
    stable/metastable energy is changing.  (Energy is in eV on all figures.)
    }
    \label{homobarr}
\end{figure}

We measured the effect of strain on the diffusion barrier.  The
substrate lattice was compressed in the horizontal direction by a
given factor, and was allowed to relax vertically.  Then an adatom was
placed on top, and the whole system was allowed to fully relax.
Fig.~\ref{homobarr} shows the energy of the system when a Ag adatom
was placed on a stable (fcc), metastable (hcp) and bridge point of a
stressed Ag(111) substrate.  The diffusion barrier (the difference of
the bridge and the stable/metastable energy) is also plotted.

Near zero stress the barrier was close to a linear function of the lattice
constant, with increasing barrier for tensile strain. This is the expected
behavior: under compressive strain the energy landscape becomes more uniform,
while under tensile strain the adatom feels more the separate attracting
potential of the surface atoms. For large tensile strain this trend breaks
down: the surface becomes softer, bringing down bridge energies, resulting in
a smaller diffusion barrier. However, this linearity was only over
a rather restricted range of strains.
Note that we do not reproduce
the result of Ratsch \emph{et al.} \cite{Ratsch} who used the LDA, and
did find linearity of the diffusion barrier with strain, as suggested
on phenomenological grounds by Dobbs \emph{et al.}\cite{dobbs1} 
On the other hand, effective medium theory calculations\cite{brune} agree
with our results,  up to a 10~meV systematic shift, see Fig.~\ref{homobarr}a.

When the lattice is unstressed, the fcc adsorption sites are slightly lower in
energy than the hcp sites. However, in our calculations this trend reverses
for large tensile strain. Note that the major effect here is not on the
bridge energies, but on the energies of the stable sites, contrary
to the effect found by Schroeder and Wolf \cite{schroeder}.

\begin{figure} 
    \epsfxsize=8.6cm 
    \epsfbox{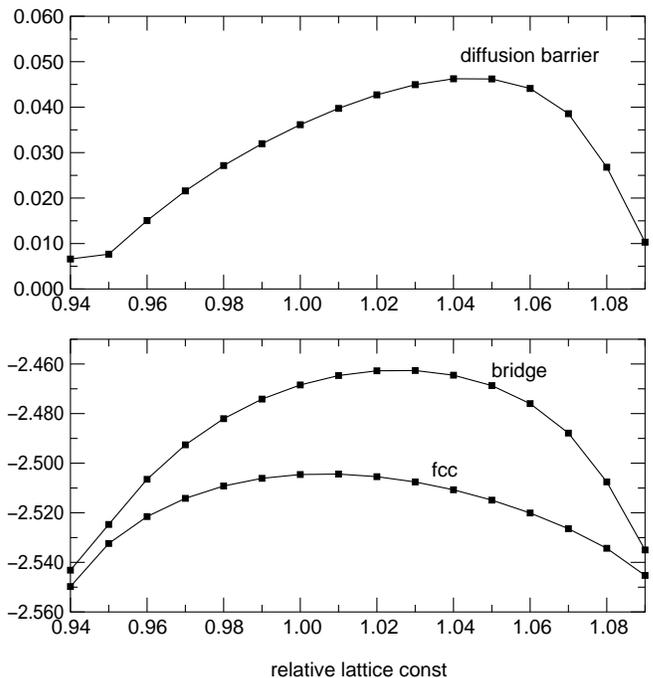}
    \bigskip
    \caption[Diffusion barrier of Ag adatom on stressed Ni(111) substrate]{
    Diffusion barrier of Ag adatom on stressed Ni(111) substrate.
    a) diffusion barrier and b) bridge and fcc adsoption site energies as
    function of 
    the ratio of the stressed and equilibrium lattice constant. This case
    the fcc site energy is constant near equilibrium, and the bridge energy is
    changing. 
    }
    \label{heterobarr}
\end{figure}

We applied the same procedure to the Ag/Ni(111) heterodiffusion system,
the barriers and energies are depicted on Fig.~\ref{heterobarr}. While the
behavior of the diffusion barrier is qualitatively the same as in the Ag
self-diffusion case, the dependence of energies on strain is different. Around
zero stress, the stable sites are unaffected, and the bridge energy
changes. From this we can draw the conclusion that whether the energy of the
stable sites or the bridge point changes under stress is system dependent, no
general statements can be made. 

\begin{figure} 
    \epsfxsize = 8.6cm
    \epsfbox{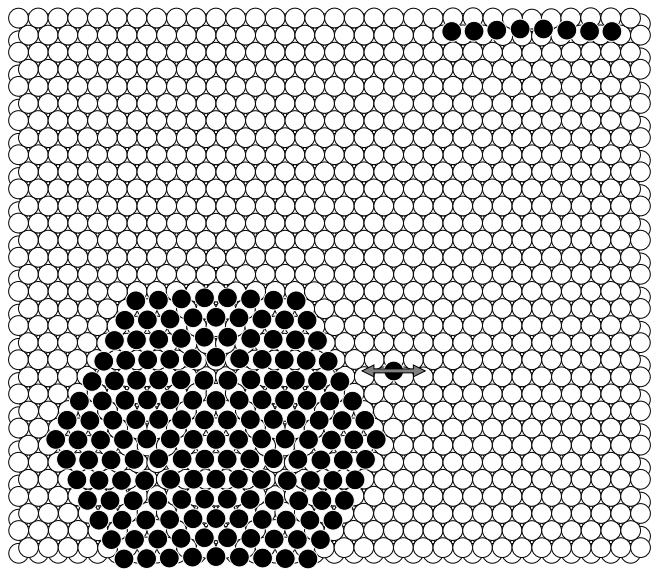}
    \bigskip
    \caption[The configuration to measure the effect of an island on the
        energy landscape]{
    The configuration to measure the effect of an island on the energy
    landscape. White circles denote substrate atoms, black ones are the hetero
    atoms. The hexagonal island is of radius 7 on this figure, the black atoms
    on the top right corner are part of the island because of the periodic
    boundary conditions. The adatom is moved in the direction of the arrow.
    }
    \label{island}
\end{figure}

One of our goals is to study the elastic effects of an island on the energy
landscape observed by the diffusing adatoms. To pursue this we deposited a
large hetero-island and an adatom on the substrate, and computed the energy of
the system for different positions of the adatom, the configuration is shown on
Fig.~\ref{island}.

\begin{figure} 
    \epsfxsize =8.6cm
    \epsfbox{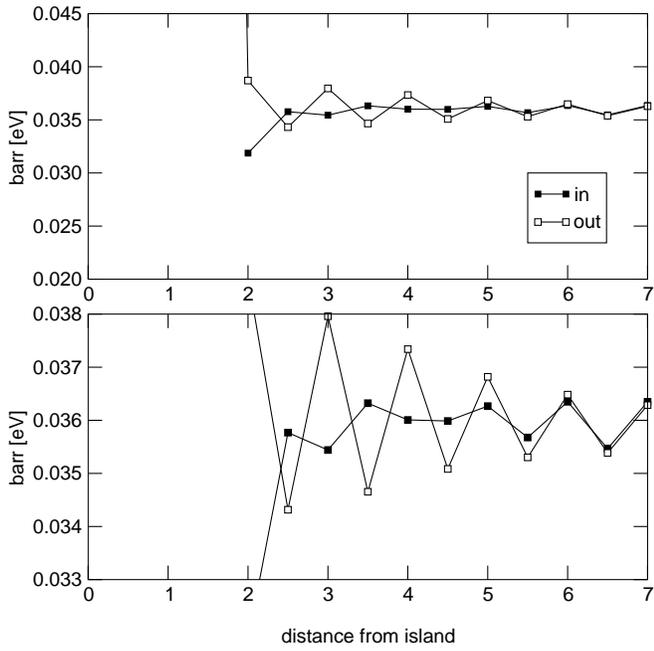}
    \bigskip
    \caption[Diffusion barrier of Ag on Ni(111) near a small Ag island]{
    Diffusion barrier of Ag on Ni(111) near a small Ag island (radius is 4
    atoms). The island is pseudomorphic. The bottom figure is magnification of
    the top figure around the equilibrium barriers.
    }
    \label{diff4}
\end{figure}

In Fig.~\ref{diff4} we plot the diffusion barriers of a Ag adatom on top of
Ni(111) substrate, as a function of the distance from a Ag island of radius
of 4 atoms. 
There are two different barriers: one seen by an adatom diffusing
away from the island, and a different one for approaching it. The oscillation
is due to the nature of the lattice: on top of an fcc(111) lattice an adatom
can be in the fcc site (stable) or hcp site (metastable). The
diffusion barrier is measured between the bridge point and the stable or
metastable site.

According to the results, near the island it is easier to diffuse away from a
stable site, and easier to diffuse inward from a metastable site. The island
does not have a strong attractive or repulsive long-range effect on the
adatom. However if the adatom is very close, it can only diffuse inwards: it
is captured by the island.

\begin{figure} 
    \epsfxsize = 8.6cm
    \epsfbox{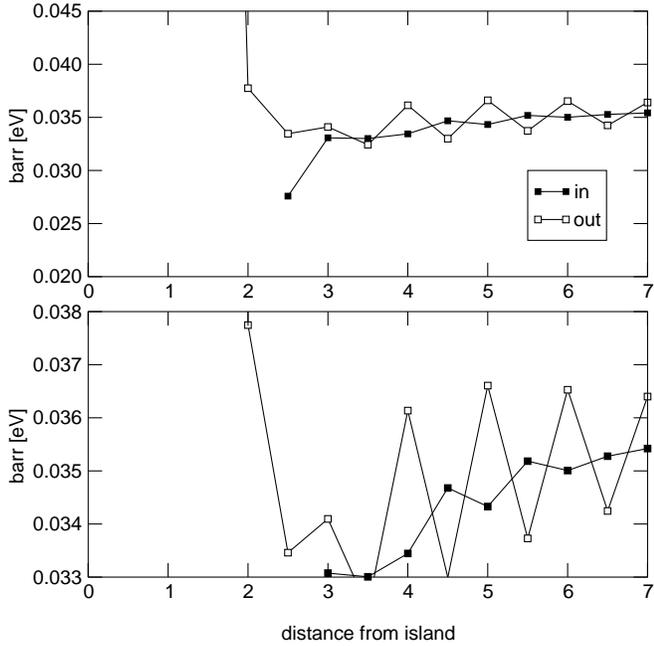}
    \bigskip
    \caption[Diffusion barrier of Ag on Ni(111) near a large Ag island]{
    Diffusion barrier of Ag on Ni(111) near a large Ag island (radius is 7
    atoms as in Fig.~\ref{island}).  The island is not pseudomorphic. The
    scale of the plots is the same as on the previous figure.
    }
    \label{diff7}
\end{figure}

The small island of the previous result was pseudomorphic with the substrate.
For larger islands this is not the case. Fig.~\ref{diff7} shows the diffusion
barriers near an island of radius of 7 atoms, which is already not
pseudomorphic,
as can be seen on Fig.~\ref{island}. The distortion of the energy landscape is
much larger in this case, and the attraction of the island can be felt at
larger distances. The effect of the island is not only attraction (the outward
barriers larger than the inward ones) but also enhancing diffusion near the
island: the diffusion barriers in both directions are decreased. Probably this
is due to the fact that the substrate near the compressed island is also
compressed.

\begin{figure} 
    \epsfxsize = 8.6cm
    \epsfbox{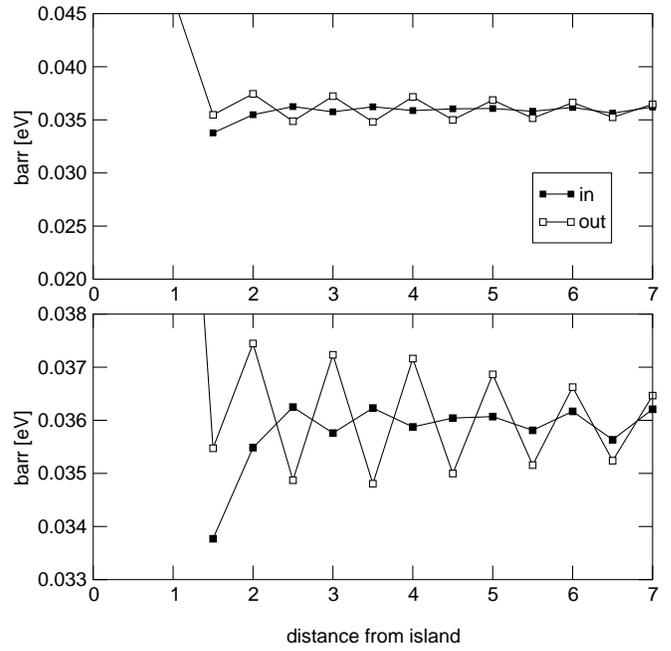}
    \bigskip
    \caption[Diffusion barrier of Ag on Ni(111) near a large Ni island]{
    Diffusion barrier of Ag on Ni(111) near a large Ni island (radius is 7
    atoms).  Same scale as previous figure.
    }
    \label{diffhomo}
\end{figure}

To check that how much of this effect is due to the presence of the compressed
hetero-island, we repeated the previous calculation with homoepitaxial island:
the large Ag island has been replaced with same size Ni island. The obtained
barriers (Fig.~\ref{diffhomo}) show even smaller effect than the case of the
small hetero-island. On a considerable range the energy landscape is deformed:
the outward and inward directions are not equivalent (as in a sawtooth
potential) but there is no global attraction or repulsion.

\begin{figure} 
    \epsfxsize = 8.6cm
    \epsfbox{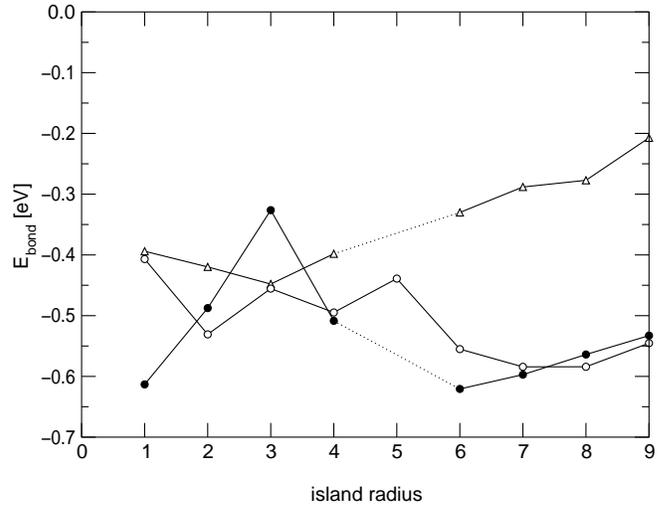}
    \bigskip
    \caption[Bonding energy to strained islands]{
    Bonding energy of an atom to strained Ag islands as a function of island
    radius. The three curves represent different positions of the bonding atom:
    an extra atom on the middle of the edge of a hexagonal island (triangles),
    the corner atom of the island (full circles), and the neigbor of a corner
    atom where the corner atom already absent (empty circles). The case of
    radius=5 is explained in the text.
    }
    \label{detach}
\end{figure}

\begin{figure} 
    \epsfxsize = 8.6cm
    \epsfbox{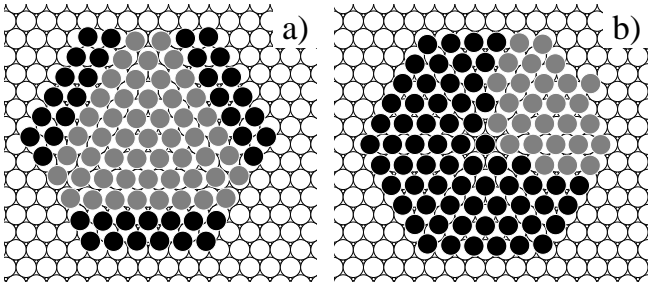}
    \bigskip
    \caption[Pseudomorphic island]{
    Relaxed island of radius 5 without and with an adjacent adatom.
    The ``pseudomorphic atoms'' are grey.
    (An atom is considered pseudomorphic if it is closer to the stable site
    extrapolated from the lattice than to other stable or metastable sites.)
    a) Without adatom: The majority of the island is pseudomorphic, only the
    edges are pushed out. Note the deformed edges.
    b) With adatom: The perturbation of the adatom was enough that only the
    nearby part of the island is pseudomorphic.  The other parts are also
    relaxed, with smooth dislocation network connecting the relaxed parts.
    }
    \label{detachii}
\end{figure}

We measured the detachment barrier from a strained island. 
Fig.~\ref{detach} shows the binding energy as a function of island size, it is
the same as the detachment barrier up to a sign.  
The trend is decreasing barrier for large enough islands in all cases, as
expected.
For large islands the detachment barrier of an extra atom at the middle of the
hexagonal island's edge (see Fig.~\ref{detachii}b) is smaller than the
detachment barrier of the corner atom, or the next atom after the corner.
This is plausible, as the extra atom at the middle of the edge is less
coordinated than the compared atoms.

It has to be noted that the binding energy of the island of radius of 5 is very
different compared to the nearby sizes. The explanation is the following. The
binding energy is defined as the energy of the island with an adjacent adatom,
the zero point is when the adatom is infinitely far away. The island of this
size is at the borderline of pseudomorphic and not pseudomorphic islands. When
we measured the energy of the island in itself, the relaxation converged to a
pseudomorphic state, see Fig.~\ref{detachii}a. But when the adatom was added,
this was enough perturbation that the system converged to a not pseudomorphic
state (Fig.~\ref{detachii}b). Thus the addition of the adatom triggered a much
lower energy state, hence the large negative bonding energy. It is possible
that the bare island also has a lower energy non-pseudomorphic state, but we
did not do a detailed search.

\begin{figure} 
    \epsfxsize = 8.6cm
    \epsfbox{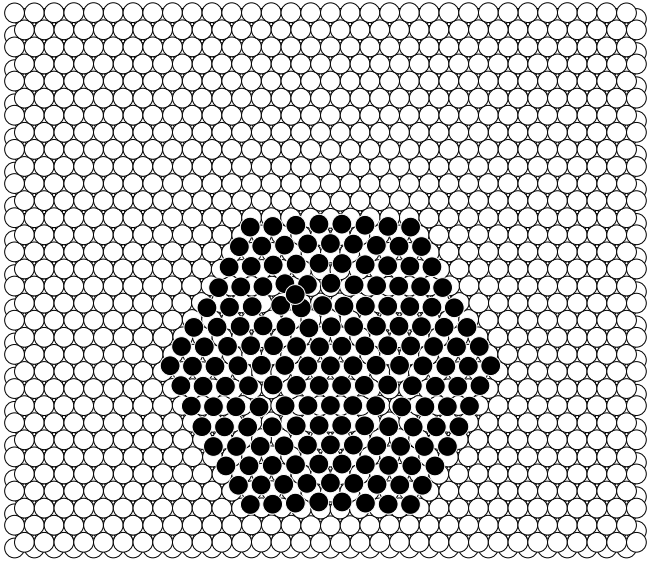}
    \bigskip
    \caption[Four-fold hollow site on an island]{
    Unusual deformations like this four-fold hollow site occur on top of an
    island.
    }
    \label{fourfold}
\end{figure}

We also tried to obtain an energy landscape on top of an island. This was
quite difficult, because the island atoms are very soft, deform very much in
the presence of an adatom on top, and there is no well defined stable,
metastable and bridge site. Fig.~\ref{fourfold} depicts a case when a the
adatom is in a deformed four-fold hollow site.

\section{Summary}
In this paper we studied the elastic effects of heteroepitaxial islands on
diffusion using atomistic simulations with EAM potential. Compressive strain
enhances diffusion, small tensile strain hinders it, but large tensile strain
also tends to enhance it. Whether the energy of the stable site changes or the
bridge energy, depends on the system.

The energy landscape near a compressed island is deformed: the island
attracts the adatom, and the diffusion is increased near the
island. Even a homoepitaxial island deforms the energy landscape, but
the change is much smaller, and only the
symmetry of the potential is broken.

The detachment barrier from a compressed island decreases with larger island
size. The diffusion barriers on top of an island are hard to measure, because
the island is soft and distorted near an adatom, there is no well defined
diffusion path. This is probably due to the fact that we chose to
work with a system that dislocates easily.

Our general conclusion from this detailed microscopic study is, in some
sense, negative. Empirical theories depend on making general statements
about the effects of strain which can be modeled with a few parameters.
Our study shows that while the qualitative ideas behind these theories
are correct -- a large two-dimensional island is destabilized by strain,
for example -- the form of the effect is quite complicated. Also, the
representation of the diffusion barriers as linear in the strain is
true only over a limited range in our calculations, and in the 
EMT \cite{brune}, while the LDA \cite{Ratsch} does give linearity.

The complexity of our results may be due to the small sizes of the
two-dimensional islands that we were able to deal with, and to
the fact that our metallic systems dislocate. Still, we think
that these results should serve as a warning against a naive 
application of continuum elasticity theory in this area. 
We should note
that in Refs. \onlinecite{Williams,Zangwill} 
unreasonable assumptions about the size of elastic effects 
were found to be necessary: Elastic couplings would have to be much larger 
than any effect that we calculate here in order to
give  significant narrowing. 
In our opinion the physical reason for the narrow size distribution
of quantum dots is still obscure.

We are grateful to Brad Orr for useful discussions.
This work is supported by DOE grant DEFG-02-95ER-45546.

\end{document}